\documentclass[twocolumn,showpacs,showkeys,pra]{revtex4-2}
\usepackage{amsmath}
\usepackage{graphicx}
\usepackage{color}
\usepackage{amssymb}
\usepackage{float}

\begin{document}
\title{	
Josephson-type oscillations in spin-orbit coupled Bose-Einstein condensates with nonlinear optical lattices}
\author{Sumaita Sultana}
\affiliation{Department of Physics, Kazi Nazrul University, Asansol-713340, W.B., India}
\author{Golam Ali Sekh}
\email{skgolamali@gmail.com}
\affiliation{Department of Physics, Kazi Nazrul University, Asansol-713340, W.B., India}
\begin{abstract}
We consider spin-orbit coupled Bose Einstein Condensate in presence of linear and nonlinear optical lattices within the framework of quasi-one-dimensional Gross-Pitaevskii equation. The population imbalance between the states changes periodically with time and the oscillation amplitude depends sensitively on the initial phase. The optical lattice is found to change phase velocity of the oscillation of population imbalance. This oscillation can also be arrested  beyond critical values of parameters. We find that the optical lattice can efficiently be used to control the critical point. 
\end{abstract}
\maketitle
\section{Introduction}
Since the first experimental observation of BEC in 1995\cite{r1,r2}, it has been playing a very important role both theoretically and experimentally for  studying various topics related to many- body system, condensed matter and nonlinear physics. Ultracold atomic system serves as a good platform, primarily,  because of it's flexibility to tune different parameters. One of the greatest advance in this  filed is the study of charge physics using neutral atoms. This leads to the realization of spin-orbit coupling in BECs \cite{r3,r4}.

Mean-field interaction in Bose-Einstein condensates introduces nonlinearity. Experimental flexibility to control interaction with the help of Feshbach resonance technique  allows  the generation of matter-wave bright and dark solitons \cite{r5,r6}. Introduction of artificial lattice potential leads to the production of another type of soliton called matter-wave gap soliton in the BECs\cite{r7}.

Experimental ability to load BEC in optical lattices immediately  grabs the attention to explore it in studying condensed matter physics \cite{r8,r8a}. The main advantage lies in the possibility to manipulate  lattice depth and spacing  just by controlling intensity of laser beams  and the angle between beams at which they are imposed on the condensate. One can induce periodicity in the condensate also by  spatial periodic modulation of scattering length by optical means.  This is the so-called nonlinear optical lattice (NOL). The NOL produces narrow state by squeezing the density distribution and it is found very efficient to control dynamics of matter-wave soliton\cite{r9,r9a}.

The possibility to realize matter-wave in SOC-BEC \cite{r10, r11} has motivated to revisit several  interesting concept of quantum mechanics \cite{r4,r12,r13,r14}. Josephson oscillation is one of them. It is a tunneling phenomenon which is generally seen in superconducting materials. In SOC-BEC one can realize a similar phenomenon\cite{r8}, often called,  Josephson-type (JT) oscillation where atomic imbalance between the components of a BEC is found to execute oscillatory behaviour if initial phase difference is induced between them \cite{r15,r16}. The oscillation of population imbalance stops  beyond the critical value of a properly chosen parameter.  This is the so-called macroscopic quantum self-trapping(QMST)\cite{r23}. 

Our objective is to study the dynamics of atomic population imbalance between the SOC components in linear and nonlinear optical lattices.  We see that the OLs induces a temporal phase in the JT oscillation and thus  causes the oscillation  to precede over  the same in absence of OLs with the progress of time. This oscillation can however be stopped by properly tuning the nonlinear optical lattice. In the recent past, Abdullaev et al. considered SOC-BEC with time-dependent Rabi frequency  in absence of OLs and studied the possibility of JT oscillation\cite{r16}.  The Josephson physics is also investigated by Garcia-March et al by two-mode approximation in elongated SOC-BECs \cite{r16a}.  Recently, the  effects of quantum  fluctuation on QMST have been studied using Lee-Huang-Yang term by Abdulleav et al\cite{r16b}.

We present a mathematical formulation based on variational approach in section II. In section III, we discuss about linear energy spectrum in presence of OLs and also find effective potential for the center of mass frame. In section IV, we study change of population imbalance  for different phases and discuss the effects of optical lattices. We also derive a condition for quantum mechanical self-trapping (QMST). In section V, we present the results obtained from numerical simulation of GP equation with a view to justify the variational results. Finally, we make  concluding remarks in section VI.
\section{General Formalism}
We consider  spin-orbit coupled Bose-Einstein condensates in quasi-one-dimensional (Q1D) harmonic trap in presence of optical lattices.  This system can appropriately be described by the following Gross-Pitaevskii equation(GPE)\cite{r16}. 
\begin{eqnarray}
i\frac{\partial \chi}{\partial t}
&=&\biggl[-\frac{1}{2}\frac{\partial^2}{\partial x^2}+ik_s \sigma_z\frac{\partial}{\partial x}+V_{ext}(x)+ \Omega(t)\sigma_x\biggl]\chi
\nonumber\\ &-&\left(\begin{array}{cc}
\gamma|\psi_a|^2+\beta|\psi_b|^2 & 0 \\
0 & \beta|\psi_a|^2+\gamma|\psi_b|^2
\end{array}\right)\chi.
\label{eq1}
\end{eqnarray}
with $\chi=(\psi_a,\psi_b)^T $. Here $\psi_a(x,t)$ and $\psi_b(x,t)$ represent the order parameters of the first and second pseudo-spin state respectively. The constants $\gamma$ and $\beta$ give \textcolor{blue}{strengths of nonlinear inter- and intra-component interactions} and $k_s$, the spin-orbit coupling strength. The symbols $\sigma_z $ and $ \sigma_x $ stand for the Pauli spin matrices.  

In Eq. (\ref{eq1}), the external potential $V_{ext}$ is given by
\begin{eqnarray}
V_{ext}=\frac{1}{2}{({\omega_x}/{\omega_\perp})^2 }x^2+V_0 \cos(2k_l x).
\label{eq2}
\end{eqnarray}
Here the first term gives the harmonic trapping potential and second term is the  linear optical lattice(LOL) with wave number $k_l$ and strength $V_0$. We  measure energy, length and time of the system in the units of  $\hbar \omega_\perp $, $a_\perp=\sqrt{{\hbar}/{m\omega_\perp}}$, $({\omega_\perp})^{-1}$ respectively and thus rewrite Eq.{\ref{eq1}} in dimensionless units. Understandably, effects of the first term in Eq. (\ref{eq2}) can be negligible if $\omega_\perp \gg \omega_x$. The nonlinear optical lattices (NOLs)  which introduce  periodicity in the system by modulating  mean-field interaction terms are given by \cite{r17,r18}
\begin{subequations}
\begin{eqnarray}
\gamma=\gamma_0+\gamma_1 \cos(2k_n x),
\label{eq3a}
\end{eqnarray}
\vskip -0.5cm
\begin{eqnarray}
\beta=\beta_0+\beta_1 \cos(2k_n x),
\label{eq3b}
\end{eqnarray}
\end{subequations}
where $k_n$ is the wave number of the NOLs with their strengths $\gamma_1$ and $\beta_1$ respectively. Time varying Raman frequency $\Omega(t)$  in Eq. (\ref{eq1}) is given by
\begin{equation}
\Omega(t)=\Omega_0+\Omega_1 \cos(\omega t).
\end{equation}
Here $\Omega_1$ and $\omega$ stand for the amplitude and frequency of $\Omega(t)$.

Eq. (\ref{eq1}) supports matter-wave solitons for $k_s \rightarrow 0$ and $\Omega(t) \rightarrow 0$. In the recent past, it is shown that SOC-BEC can support matter wave solitons\cite{r11,r19}. However, the nature of soliton depends on the relative values of $\Omega_0$ and $k_s$. \textcolor{blue}{In this work we consider the case  $ k_s^2<\Omega_0$ where it permits bright solitons solution.}  Thus, keeping the similarity with shape of solution, we adopt
\begin{equation}
\left(\begin{array}{cc}
\psi_a\\
\psi_b
\end{array}\right)=\left(\begin{array}{cc}
\textcolor{blue}{(\frac{N_a}{\sqrt{\pi} w})^{1/2}}\,\, e^{-(x-x_0)^2/2w^2+ik_a (x-x_0)+i\phi_a}\\
\textcolor{blue}{(\frac{N_b}{\sqrt{\pi} w})^{1/2}}\,\, e^{-(x-x_0)^2/2w^2+ik_b (x-x_0)+i\phi_b}
\end{array}\right)
\label{eq5}
\end{equation}
as trial solutions of the system to  work within the framework of variational approach. \textcolor{blue}{Understandably, the norm  $N=\int\left(|\psi_a|^2+|\psi_b|^2\right)\,dx=(N_a+N_b)$}. Here $N_j$, $x_0 $, $w $, $k_j $ and $ \phi_j $($j=a, b$) are time dependent variational parameters which represent respectively \textcolor{blue}{number of particles}, center of mass, width, wave number and phase respectively. \textcolor{blue}{The variational ansatzs in Eq.(\ref{eq1}) are so chosen  that they represent spatially overlapped spin-up and spin-down solitons since, for SU(2) atomic interaction, the two-component solitons prefer mixed phase \cite{r20,r21}.}  
 
We obtain the following Lagrangian density from Eq.(\ref{eq1}) using inverse variational approach.
\begin{eqnarray}
\mathcal{L}&=&\bigg[\frac{i}{2}\bigg(\psi_a^*\frac {d\psi_a}{dt}+\psi_b^*\frac {d\psi_b}{dt}\bigg)\nonumber\\&-&\frac{ik_s}{2}\bigg(\psi_a^*\frac{d\psi_a}{dx}-\psi_b^*\frac{d\psi_b}{dx}\bigg)+c.c\bigg]\nonumber\\&-&\frac{1}{2}\bigg|\frac{d\psi_a}{dx}\bigg|^2-\frac{1}{2}\bigg|\frac{d\psi_b}{dx}\bigg|^2-\Omega \psi_a^* \psi_b-\Omega \psi_b^* \psi_a+\frac{\gamma_0}{2}|\psi_a|^4
\nonumber\\&+&\frac{{\gamma_0}}{2}|\psi_b|^4+{\beta_0}|\psi_a|^2|\psi_b|^2 
-V_0 \cos(2k_l x)|\psi_a|^2\nonumber\\&-&V_0 \cos(2k_l x)|\psi_b|^2+\frac{\gamma_1}{2}\cos(2k_n x)|\psi_a|^4\nonumber\\&+&\frac{\gamma_1}{2}\cos(2k_n x)|\psi_b|^4+\beta_1 \cos(2 k_n x)|\psi_b|^2|\psi_a|^2.
\label{eq6}
\end{eqnarray}
Substituting Eq.(\ref{eq5}) in Eq. (\ref{eq6}) and then integrating with respect to $x$ from $-\infty$ to $+\infty$ we get the following  averaged Lagrangian density.
\begin{eqnarray}
L=&-&\sum_{j\in \{a,b\}} N_j\bigg[\frac{d\phi_j}{dt}-k_j\frac{dx_0}{dt}+\frac{1}{4w^2}+\frac {k_j^2}{2}\bigg]\nonumber\\&-&2\Omega(t)\sqrt{N_aN_b}e^{-w^2k_-^2}\cos(\varphi)+(k_aN_a-k_b N_b)k_s \nonumber\\&+&\frac{1}{2\sqrt{2\pi }w}(\gamma_0 N_a^2+\gamma_0 N_b^2+2\beta_0 N_a N_b)
\nonumber\\&+&\frac{e^{\frac{-k_n^2 w^2}{2}}}{2\sqrt{2\pi}w} (\gamma_1 N_a^2+\gamma_1 N_b^2+2\beta_1 N_a N_b)\cos(2 k_n x_0)\nonumber\\&- &V_0 (N_a+N_b)\,e^{-k_l^2 w^2} \cos(2 k_l x_0).
\label{eq7}
\end{eqnarray}
Here $ \varphi=\phi_a-\phi_b $. Making use of the Ritz  optimization conditions: $\frac{\delta L}{\delta w}=0$, $\frac{\delta L}{\delta N_a}=0$, $\frac{\delta L}{\delta N_b}=0$, $\frac{\delta L}{\delta \phi_a}=0$, $\frac{\delta L}{\delta \phi_b}=0$, $\frac{\delta L}{\delta k_a}=0$ and $\frac{\delta L}{\delta k_b}=0$,
we obtain the following coupled equations.
\begin{eqnarray}
\frac{dx_0}{dt}&=&k_+,
\label{eq8}\\
\frac{dk_+}{dt}&=&2k_s\Omega(t)\sqrt{1-Z^2}e^{-k_s^2 w^2} \sin(\varphi)\nonumber\\&-&\frac{N k_n \sin(2k_n x_0)}{2\sqrt{2\pi}w} e^{\frac{-k_n^2 w^2}{2}}[\Gamma_\gamma+\Gamma_\beta Z^2]
\nonumber\\&+&2V_0k_le^{-k_l^2w^2}\sin(2k_lx_0),
\label{eq9}\\
\frac{dZ}{dt}&=&-2e^{-k_s^2 w^2}\Omega(t)\sqrt{1-Z^2}\sin(\varphi),\\
\frac{d\varphi}{dt}&=& 2k_s k_+ +\Lambda Z+\frac{2e^{-k_s^2w^2}\Omega(t)Z}{\sqrt{1-Z^2}}\cos(\varphi)\nonumber\\&+&\frac{NZ}{\sqrt{2\pi}w}e^{\frac{-k_n^2\omega^2}{2}}\cos(2k_n x_0)\Gamma_\beta.
\label{eq11}
\end{eqnarray}
Here  $ \Lambda=N(\gamma_0-\beta_0)/(\sqrt{2\pi})w$, $k_\pm=(k_a\pm k_b)/2$, $\Gamma_\beta=\gamma_1-\beta_1$, $\Gamma_\gamma=\gamma_1+\beta_1$, and choose $k_-\approx k_s$ for  weak spin-orbit coupling. \textcolor{blue}{Understandably, $ Z=(N_a-N_b)/N$ can be treated as population imbalance between the components. The  rate of change of $Z$ depends  directly on lattice parameter.} Therefore, it is essential to  solve the coupled Eqs.(\ref{eq8})-(\ref{eq11}) numerically  in order to see how the population imbalance is influenced by the different parameters of the system.

\section{Linear spectrum and effective potential of the coupled systems}
\begin{figure}[h!]
\includegraphics[scale=.25]{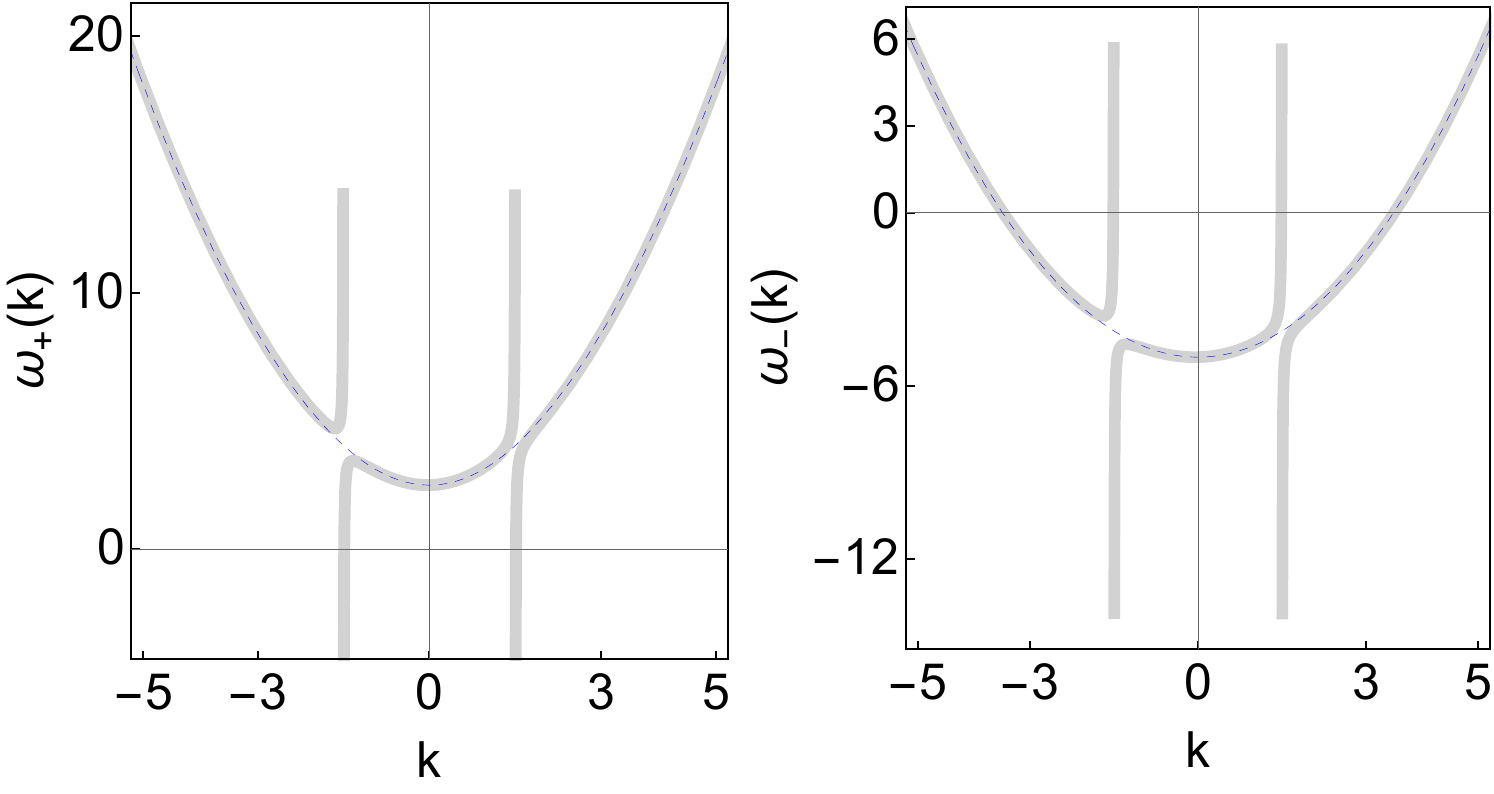}
\caption{The variation of $\omega_{\pm}(k)$ with $k$ for  $V_0=0$ (dashed blue line) and $V_0=-1$, $k_l=2$ (thick grey line). In both panel we take $\Omega_0=10$ and $k_s=2$. Here $\omega_{\pm}(k)$ and $k$ are measured in the units of $k_s^2$ and $k_s$ respectively. }
\label{fig1}
\end{figure}

We have noted that the relative values of $k_s$ and $\Omega_0$ play a crucial role for the prediction of various solutions in the system. In presence of optical lattices one can expect interplay among lattice and SOC parameters. For better understanding of the nature of solutions, we obtain the following energy-momentum relation using plane wave solution, $(\psi_a, \psi_b)=(u_0,v_0)\exp[ik x-i\omega(k)t]$,  
\begin{equation}
\omega_\pm(k)=\bigg[\frac{1}{2}k^2+\textcolor{blue}{\sqrt{\frac{2}{\pi}
}\frac{V_0k}{k^2-k_l^2}\bigg]}\pm\sqrt{k_s^2k^2+\Omega_0^2}.
\label{eq12}
\end{equation}
Here, $k$ is the wave number and $\omega$ is the frequency of the plane wave. We obtain the contribution of LOL in the linear dispersion relation by taking Fourier sine transform of  
$V_0 \cos( {2}k_l x)$ \textcolor{blue}{and then using the concept of Wigner-Seitz cell for construction of Brillouin zone\cite{r22}}. From the variation of $\omega_\pm $ with $k$ (Fig.\ref{fig1}) for  $k_s^2<\Omega_0$ see that there are two distinct  branches, namely, upper branch ($\omega_+$, left panel) and lower branch ($\omega_-$, right panel). Interestingly,  both the dispersion curves have single minimum.  In presence of optical lattice  discontinuities induce in the dispersion curve and the locations of these discontinuities depend on the  wave number ($k_l$) of the lattice. For  wave number $k$, the discontinuities appear at  $k=\pm k_l$.  Therefore, by choosing  appropriate value of $k_l$ one can expect interesting dynamical behavior of the system.  

Dynamical behavior of solution in the center of  mass frame of the coupled system is described by Eq.(\ref{eq8}).  In analogy with $m\ddot{x}_0=-d V_{\rm eff}/dx_0$, Eq. (\ref{eq8}) can  be used to find the following effective potential for a particular time.
\begin{figure}[h!]
\includegraphics[scale=.45]{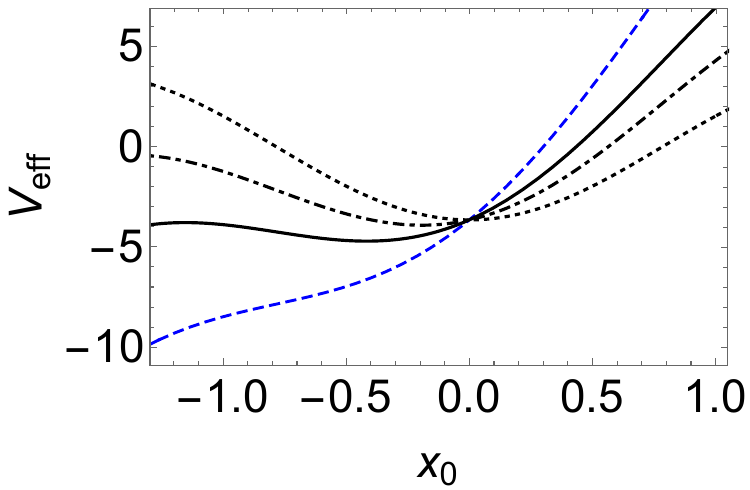}
\caption{Effective potential of  the center of mass of a soliton pairs for different values of initial phase $\varphi(0)=\pi/4$ (dashed curve, $\pi/8$(solid), $\pi/16$ (dot-dashed) and $0$ or $\pi$. Other parameters are fixed as:   $k_n=1$, $V_0=-2$ and  $\Gamma_\gamma=2.15$ and $\Gamma_\beta=0.35$, $Z(0)=0.5$ and $w=0.45$.}
\label{fig2}
\end{figure}
\begin{eqnarray}
V_{\rm eff}&=&-\frac{N}{4w\sqrt{2\pi}}e^{-k_n^2 w^2/2}\cos(2k_nx_0)[\Gamma_\gamma+\Gamma_\beta Z^2]\nonumber\\&+&V_0e^{-k_l^2w^2}\cos(2k_lx_0)\nonumber\\&-&2\Omega k_s\sqrt{1-Z^2} x_0e^{-k_s^2w^2}\sin(\varphi).
\label{eq13}
\end{eqnarray}
 Understandably, $V_{\rm eff}$ describes effective potential that can cause the soliton to move(Fig.\ref{fig2}).  In absence of optical lattice the effective potential has no minimum. It develops  local minima in presence of OLs. Depth of a local minimum increases with the decrease of initial phase difference $\varphi(0)$. We see that the effective potential becomes  periodic  with prominent local minima if $\varphi(0)$ is either $0$ or $\pi$( dotted curve). The  nature of the effective potential sensitively depends on the soliton width ($w$) and lattice parameters. Thus one can control the motion of center of mass by taking appropriate lattice parameters.
\section{JOSEPHSON-TYPE OSCILLATION BETWEEN TWO-WEAKLY COUPLED CONDENSATES}
First we consider the case where center of mass of the system is located at a fixed position in the effective potential but  the population imbalance  occurs between the components due to interaction.  For a fixed value of $x_0$, the coupled equations are reduced to the following pair of equations.
\begin{equation}
\frac{dZ}{dt}=-A\sqrt{1-Z^2}\sin( \varphi),
\label{eq14}
\end{equation}
\begin{eqnarray}
\frac{d\varphi}{dt}&=&(\Lambda+B) Z+\frac{A Z}{\sqrt{1-Z^2}}\cos(\varphi).
\label{eq15}
\end{eqnarray}
Here $A=2e^{-k_s^2 w^2}\Omega(t)$ and $B=\frac{N \,\Gamma_\beta \, e^{\frac{-k_n^2w^2}{2}}}{\sqrt{2\pi}w}\cos(2k_n x_0)$ with $\Gamma_\beta=(\gamma_1-\beta_1)$. We see that the population imbalance parameter does not directly depend on the nonlinear lattice parameter but it may depend on lattice through the phase. However, the phase becomes independent of NOL if $\Gamma_\beta=0$. Therefore, the effects of lattice can be  studied by taking appropriate values of $\Gamma_\beta$. 

\begin{figure}[h!]
\includegraphics[scale=0.305]{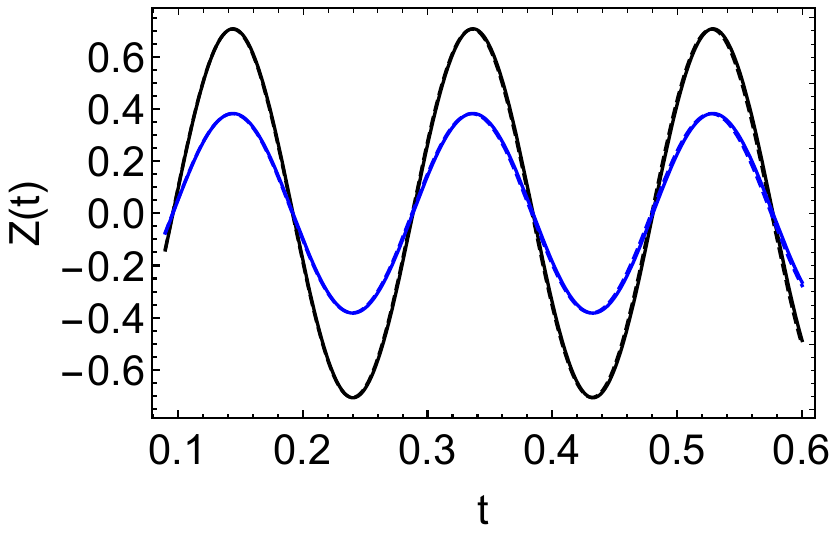}
\hskip -0.1cm
\includegraphics[scale=0.305]{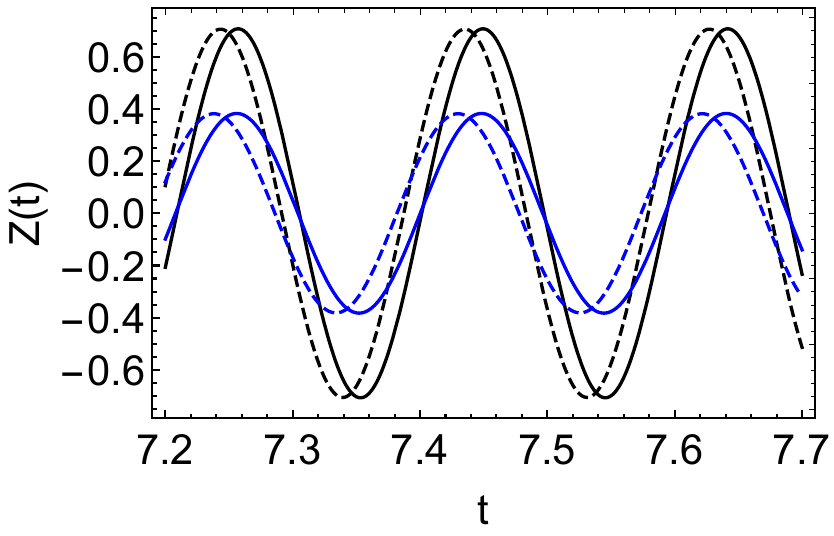}
\caption{\textcolor{blue}{Left panel:  Variation of atomic population imbalance ($Z(t)$) with $t$ for $\varphi(0)=\pi/8$(blue solid), $\varphi(0)=\pi/4$(black solid) in  presence of optical  lattices(OLs) while the dashed curves give the same in absence of OLs.  Right Panel: It gives the plots similar to those used in the left panel near $t=7$. Other parameters are fixed as: $\gamma_0=2.5$, $\beta_0=2.25$, $\Gamma_\beta=-0.25$,  $\Omega_0=20$, $ k_s=1$,  $x_0=0.25$, $w=0.45$ and $k_n=1$.}}
\label{fig3}
\end{figure}

\begin{figure}[h!]
\includegraphics[scale=.35]{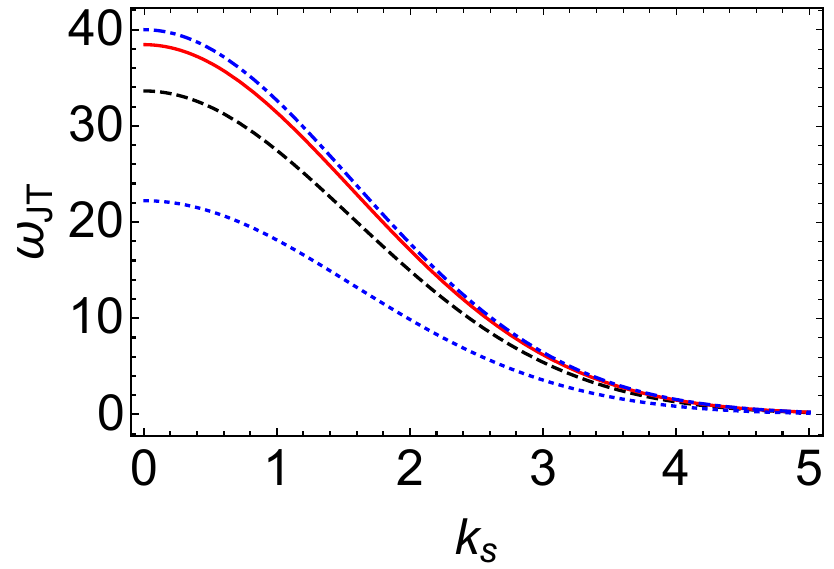}
\caption{Frequency of JT oscillation for different values of initial phases $\varphi(0)=2\pi/5$ (blue dotted), $\pi/4$ (black dashed curve), $\pi/8$(red solid), and $0$ (blue dot-dashed). Other parameters are fixed as:   $k_n=1$ and  $\Gamma_\gamma=2.15$ and $\Gamma_\beta=0.35$, $Z(0)=0.5$ and $w=0.45$.}
\label{fig3a}
\end{figure}

The variation of $Z$ with $t$ and its dependence on $\varphi$ is displayed  in Fig.\ref{fig3}. It is seen that the population imbalance oscillates with time (left panel).  Amplitude of oscillation depends on the initial phase $\varphi(0)$. The period of oscillation is independent of $\varphi(0)$ (dashed curves). The oscillation of atomic population between two SOC states resembles  with the concept of Josephson oscillation and thus we call it as Josephson-type (JT) oscillation. However, the period of oscillation is affected by the OLs (right panel, solid curve). Specifically,  OLs imprint phases and thus causes the frequency of oscillation of population imbalance to increase. The frequency of oscillation obtained from Eq. (\ref{eq14}) is given by\cite{r23}
\begin{equation}
\omega_{JT} =\left[A(\Lambda+B+\cos \varphi(0) A)\right]^{1/2}.
\label{eq16a}
\end{equation}
In writing Eq.(\ref{eq16a}) we have used $\ddot{Z}+\omega_{JT}^2 Z=0$. The oscillation frequency depends on the initial value of $\varphi(t)$ and on the values of parameters $A$, $B$ and $\Lambda$ which vary with the lattice and SOC parameters. Particularly, $\omega_{JT}$ is maximum for $\varphi(0)=0$  while it is minimum for $\varphi(0)=\pi$. Variation of $\omega_{JT}$ with different $k_s$ (Fig. \ref{fig3a}) indicates that oscillation frequencies for different $\varphi(0)$ values are distinguishable only in the limit of weak spin-orbit coupling. 

\subsection{Self-trapping of JT oscillation}
For a fixed value of initial population imbalance ($z(0)$) and phase ($\varphi(0)$), the populations of the SOC states can be macroscopically  quantum self-trapped (MQST) for a critical value of parameters. The MQST condition says that the total   energy (Hamiltonian)  $E=H[Z(t),\varphi(t)]$ of the system should satisfy the following condition\cite{r23,r24,r25}
\begin{equation}
 H[Z(0),\varphi(0)]\!=\!\frac{1}{2}(\Lambda\!+\!B)Z(0)^2\!-\!A\cos(\varphi(0))\sqrt{1\!-\!Z(\!0)^2}\!\!>\!E_-,
 \label{eq17}
\end{equation}
where $E_-$ is the value of $E$ at the
stationary point $\varphi_s=(2n+1)\pi,\,\,Z_s=0$ of Eqs.(\ref{eq14}) and (\ref{eq15}). Clearly, $E_-=A$. One can check that  the energy $(E)$ becomes $E_+=-A$ at $\varphi_s=2n\pi,\,\,Z_s=0$. In both the cases $Z-$symmetry remain valid. However, $z-$ symmetry breaks at the stationary point $\varphi_s=(2n+1)\pi,\,\,Z_s=\sqrt{1-A^2/(\Lambda+B)}$.

\begin{figure}[h!]
\includegraphics[scale=.3]{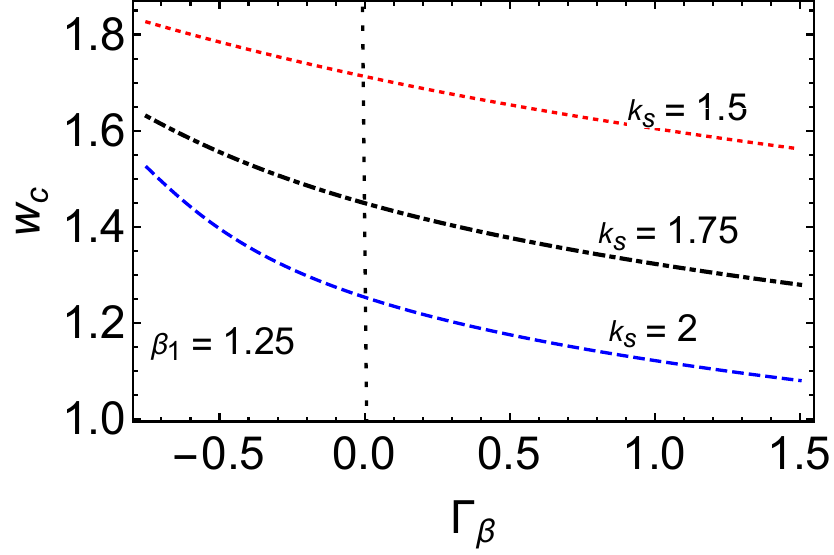}
\includegraphics[scale=.305]{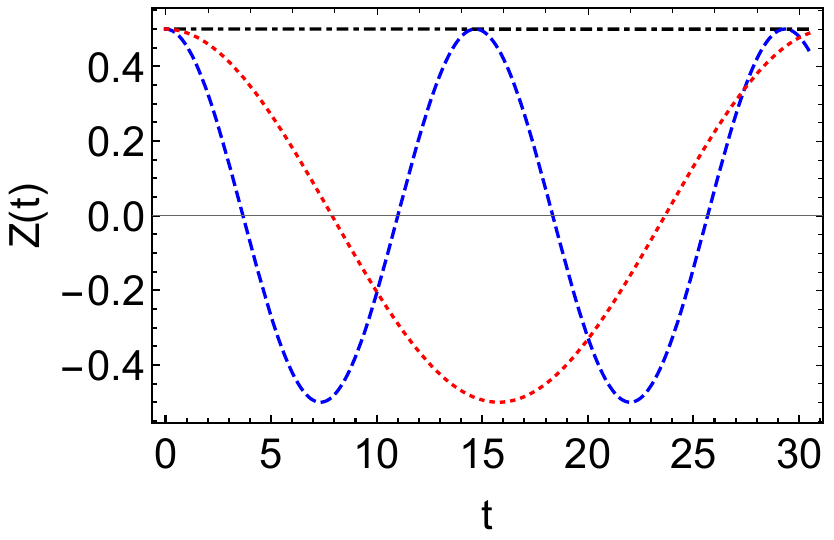}
\caption{\textcolor{blue}{Left panel: Critical width ($w_c$)  with different values of $\Gamma_\beta$ for self trapping. Here we have used $\beta_0=2.25$, $\gamma_0=2.5$, $k_n=1$, $\Omega_0=20$, $\varphi(0)=\pi$ and $Z(0)=0.5$.
Right panel: $Z(t)$ versus $t$ with $k_s=1.5$ for  $w=1.4$ (blue dashed line), $w=1.5$(red dotted line) and $w=1.66574$(black dot-dashed line).}}
\label{fig4a}
\end{figure}

 
Let us define a parameter $\Theta_c$ at the critical point of MQST as follows.
\begin{equation}
\Theta_c=(\Lambda+B)/A.
\label{eq18}
\end{equation}
Therefore, from Eq. (\ref{eq17}) we can write
\begin{equation}
\Theta_c=\frac{2}{Z(0)^2} \left(1+\sqrt{1-Z(0)^2} \cos[\varphi(0)]\right).
\label{eq19}
\end{equation}
This condition along with  Eq. (\ref{eq18}) allows us to find critical value of width($w_c$) for given values of different parameters of the system to get MQST. In Fig. \ref{fig4a}, we plot $w_c$ with $\Gamma_\beta$ for different values of $k_s$. We see that critical value of width can be greater or smaller than that in the absence of the NOL (dotted vertical line, $\Gamma_\beta=0$). For particular values of lattice strengths, critical value of width decreases with the increase of $k_s$.  The oscillation of population imbalance  below and near the critical value of width is shown in the right panel of Fig. \ref{fig4a}.   It is seen that the time period  of oscillation increases as we approach the critical width (red dotted curve). It becomes  infinitely large if width exceed the critical value. Thus, it will not possible for the atoms to change states and  they are trapped in their states (black dot-dashed curve).
\begin{figure}[h!]
\includegraphics[scale=.305]{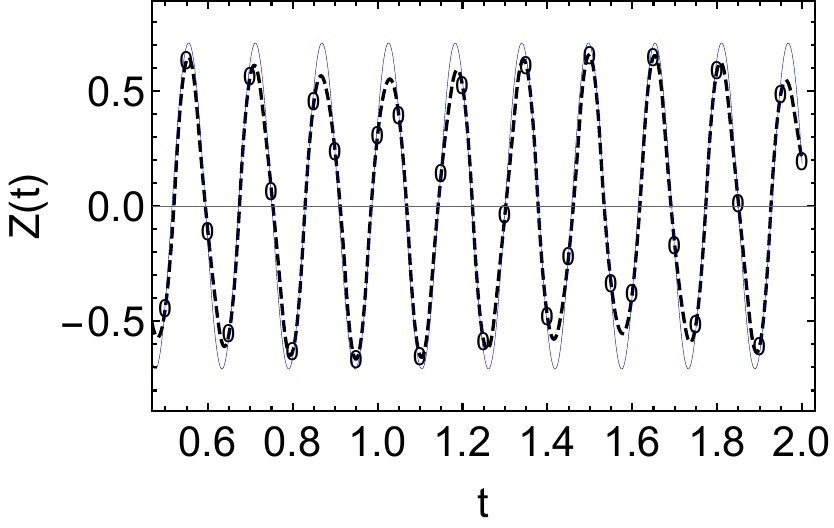}
\hskip -0.1cm
\includegraphics[scale=.305]{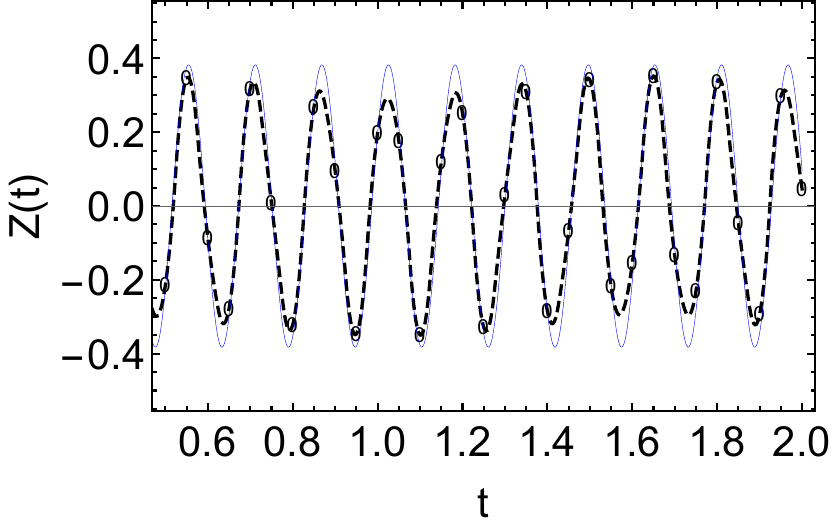}
\caption{Left panel: Numerical results of JT oscillation in SOC BEC for $\varphi(0)=\pi/4$ in presence of  nonlinear optical lattice with $\Gamma_\beta=1.15$, $\Gamma_\gamma=1.16$, $k_n=1$, $\gamma_0=2.5$, $\beta_0=2.25$, $k_s=1$, $w=0.45$, $\Omega_0=20$ and {\bf $x_0(0)=0.25$}. Right panel: It shows JT oscillation for $\varphi(0)=\pi/8$ with the same values of parameters as used in the left panel. In both the panel black dots  and solid blue curves represent the numerical and variational results respectively. }
\label{fig5a}
\end{figure}

\section{Numerical simulation of JT oscillation}
We have seen that population imbalance between spin-orbit coupled two condensates oscillates and oscillation is influenced by both SOC and lattice parameters using variational method. It is an approximation method. Therefore, it is quite essential to validate some of the obtained results through direct numerical simulation of GP equation in (\ref{eq1}).  In view of this, we use finite difference Crank-Nicholson scheme and achieve our goal using Thomas algorithm. Using Eq.(\ref{eq5}) as initial density profile, we solve Eq. (\ref{eq5}) and  we plot $Z(t)$ with $t$ obtained from direct numerical simulation of GP equation (black big dots with dashed)  and variational calculations in Fig. \ref{fig5a} for $\varphi(0)=\pi/4$(left panel) and $\varphi(0)=\pi/8$(right panel).  Clearly, numerical results shows good agreement with the variational result(solid curve).  We have also checked that $Z(t)$ versus $\varphi(t)$ describes a periodic motion.

\begin{figure}[h!]
\includegraphics[scale=.33]{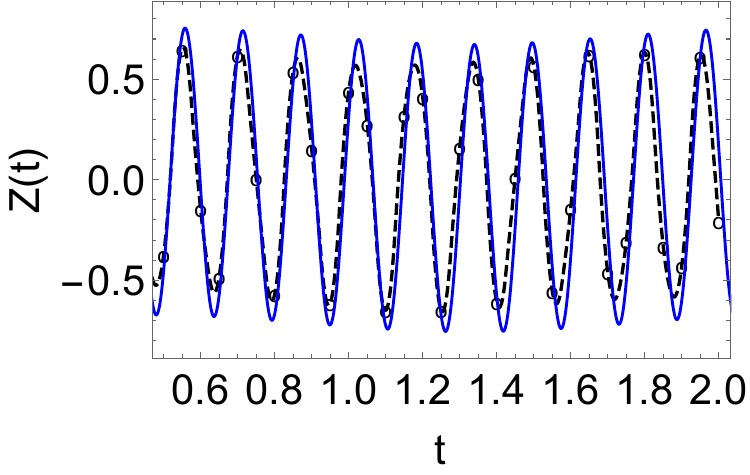}
\includegraphics[scale=.33]{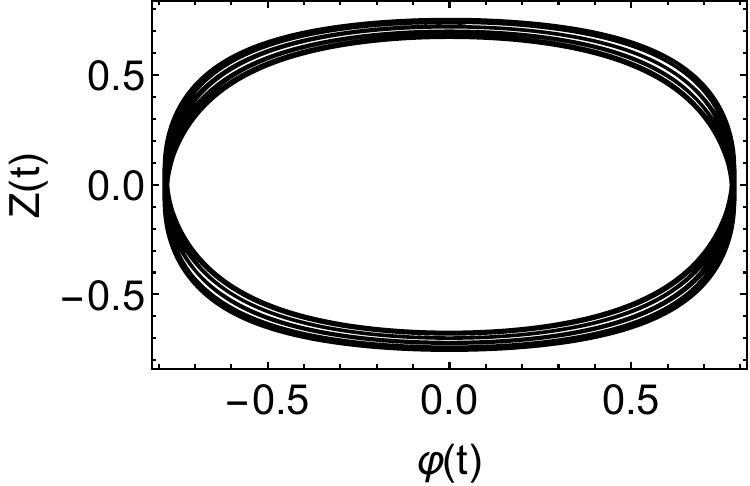}
\caption{\textcolor{blue}{Left panel: Variation of atomic population imbalance ($Z(t)$)in presence of both linear and nonlinear optical lattices for $V_0=-2$, $ \varphi(0)=\pi/4$, $\Omega_0=20$, $k_s=1$,  $\gamma_0=2.5$, $\beta_0=2.25$, $\Gamma_\beta=1.25$, $\Gamma_\gamma=1.26$ and $x_0=0.35$. We represent variational and numerical results by  solid line and dashed line with big dots respectively.
Right panel:  Variation of $Z(t)$ with $\varphi(t)$ for different values of $t$.}}
\label{fig6a} 
\end{figure}
\begin{figure}[h!]
\includegraphics[scale=.33]{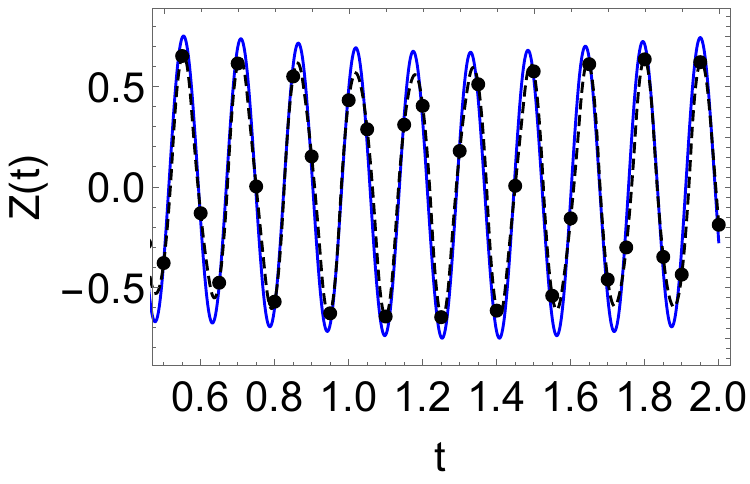}
\includegraphics[scale=.33]{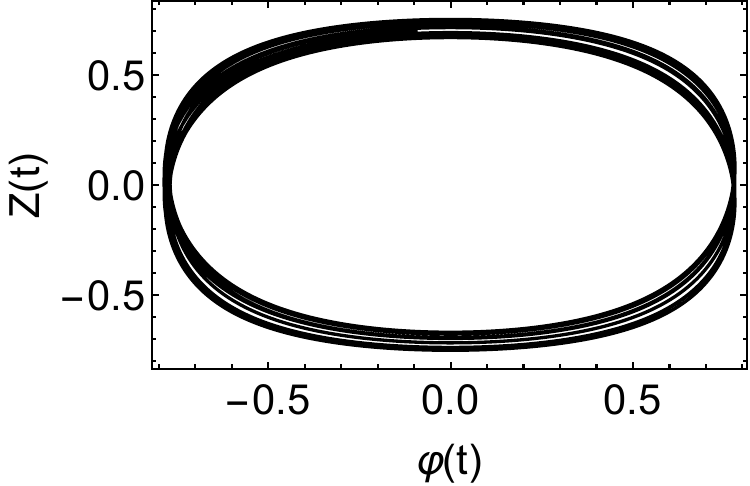}
\caption{Numerical result of JT oscillation with time-dependent Rabi frequency and OLs for $\Omega_0=20$, $\Omega_1=0.2$, $\omega=32$, $\varphi(0)=\pi/4$, $\gamma_0=2.5$, $\beta_0=2.25$, $k_s=1$ , $\Gamma_\beta=1.25$, $\Gamma_\gamma=1.26$, $V_0=-2$, $k_n=k_l=1$ and $x_0=0.35$. Right panel:  Variation of $Z(t)$ with $\varphi(t)$ for different values of $t$.}
\label{fig6}
\end{figure}
\begin{figure}[h!]
\includegraphics[scale=.21]{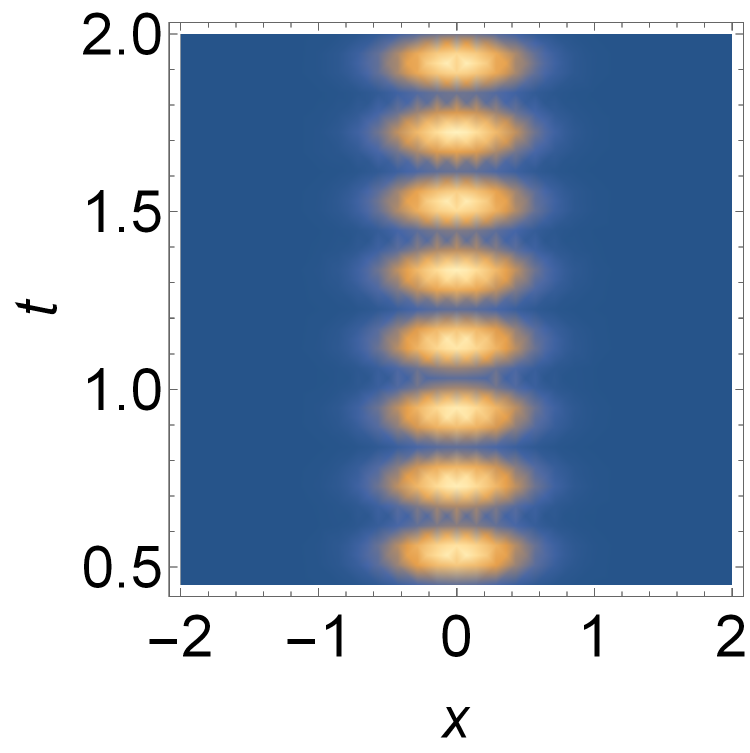}
\hspace{-0.26cm}
\includegraphics[scale=.21]{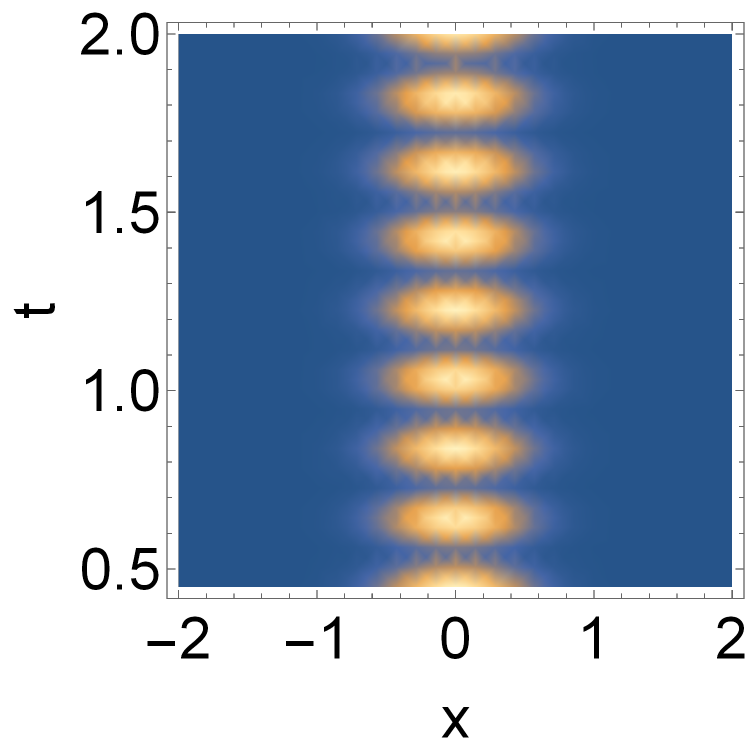}
\hspace{-0.26cm}
\includegraphics[scale=.21]{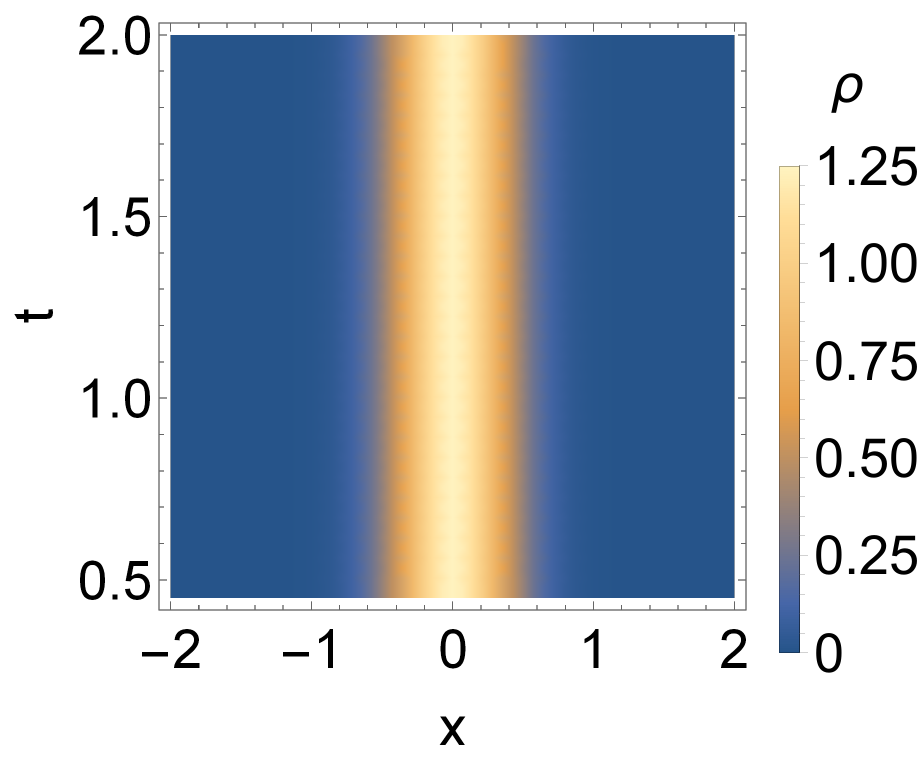}
\caption{Density plots of the first component  $\rho_a=|\psi_a|^2$ (left panel), second component $\rho_b=|\psi_b|^2$(middle panel) and total $\rho$(right panel) for $\Omega_0=20$ in presence of $\Gamma_\beta=1.25$, $\Gamma_\gamma=1.26$, $V_0=-2$, $k_n=k_l=1$, $\varphi(0)=\pi/4$, $\gamma_0=2.5$, $\beta_0=2.25$ and $k_s=1$.}
\label{fig8a}
\end{figure}

We see that  Eqs.(\ref{eq14}) and (\ref{eq15}) describing JT-oscillation do not involve the effects of linear optical lattice directly. However, one can fix initial width  of matter-wave solitons using Eq.(\ref{eq9}) from the stationary condition. Therefore, it is instructive to solve  Eqs.(\ref{eq8})-(\ref{eq11}) and compare the variational results with those obtained from  numerical simulation. The result is displayed in Fig.\ref{fig6a}. We see that atomic population difference changes periodically with slightly different frequency and the variational  result agrees the numerical result(left panel).  The $\varphi$ dependence of $Z(t)$ (right panel) shows that the dynamical response of $Z(t)$ in OLs is not exactly periodic.

In all the above cases, we have considered that the  Rabi frequency is constant. However, one can vary it with time as given in Eq.(\ref{eq5}). The variation of JT oscillation with time-dependent frequency is shown in Fig.\ref{fig5a}. We see that the envelop of the population imbalance is periodically modified due to time-dependent Rabi frequency. Here we find the numerical result serves better than variational result.

It is an interesting curiosity  to study density evolution of the spin-orbit coupled condensates during the change of population imbalance between the condensates in presence of OLs. The  plots in Fig. \ref{fig8a} clearly shows that atomic densities of both the components(left and middle panels) change periodically but the changes are  different for the two component at a particular time. For example, if the density of one component is maximum at time $t=t_1$ then the density of the other component is minimum at the same time $t_1$. However, the total density profile ($\rho$) remains constant with time (right panel).

\section{conclusion}
Bose-Einstein condensate  serves an ideal platform to study different phenomena of condensed matter physics. Josephson oscillation is the one which can be realized in spin-orbit coupled Bose-Einstein condensates (SOC-BECs) through the oscillation of population imbalance between the coupled states. We consider Josephson-type oscillation in SOC-BECs with optical lattices and discuss its dependence on lattice and SOC parameters. More specifically, oscillation amplitude depends on the initial phase difference between components. Optical lattice changes phase velocity and thus increases the frequency of JT oscillation. 

During the oscillation of the maximum value of population (MVP) both state are equal implying that average value of population imbalance is zero. However, it is also possible to create oscillation with unequal MVP in the SOC states. This is termed as quantum mechanical self-trapping(QMST). The QMST condition allows to determine critical value of width for the observation of self trapping. For given values of SOC parameters,lattice parameter can be used to control the value of critical parameter for QMST.

\section*{Acknowledgements}
S. Sultana would like to thank  "West Bengal Higher Education Department" for providing Swami Vivekananda Merit Cum Means Scholarship. GAS  would like to acknowledge the funding from the “Science and Engineering Research Board(SERB), Govt. of India" through Grant No. CRG/2019/000737.

\end{document}